# Leaking LoRa: An Evaluation of Password Leaks and Knowledge Storage in Large Language Models

Ryan Marinelli[1][0009-0001-7279-3156] and Magnus Eckhoff[1,2]

[1] University of Oslo, Department of Informatics
P.O. Box 1080 Blindern, 0316 Oslo, Norway
ryanma@ifi.uio.no, magnusec@ifi.uio.no
[2] Norwegian Defence Research Establishment (FFI)
Kjeller, Norway

**Abstract.** To effectively deploy Large Language Models (LLMs) in application-specific settings, fine-tuning techniques are applied to enhance performance on specialized tasks. This process often involves fine-tuning on user data data, which may contain sensitive information. Although not recommended, it is not uncommon for users to send passwords in messages, and fine-tuning models on this could result in passwords being leaked. In this study, a Large Language Model is fine-tuned with customer support data and passwords from the RockYou password wordlist using Low-Rank Adaptation (LoRA). Out of the first 200 passwords from the list, 37 were successfully recovered. Further, causal tracing is used to identify that password information is largely located in a few layers. Lastly, Rank One Model Editing (ROME) is used to remove the password information from the model, resulting in the number of passwords recovered going from 37 to 0.

## 1 Introduction

In an effort to stay competitive, more and more companies are utilizing Large Language Models. This can improve both the workflow of the employees and customer interactions. With LLMs being by nature large, they require both large amounts of computational resources and data to train. As a result training custom LLMs is less common in favor of fine-tuning existing models. A dataset with task specific data can be used to modify the parameters of an existing pre-trained LLM resulting in the model performing better on task-specific applications. If the data used to fine-tune contains sensitive information that should not be shared there is a risk that the information can be retrieved by threat actors. Although passwords and keys are advised against being stored in documents or other files, it is still not uncommon for end-users to do so [15]. Similarly, in customer support, it is not uncommon for an unknowing user to give away their credentials even when asked not to.

The bleeding of credentials is not only an operational issue but also an infrastructural. This was infamously problematic with Github Copilot leaking secrets from careless developers [14]. Github Copilot was trained on all public Github repositories, both large and small. Going against standard practice, some developers had hard coded API keys as strings in the code, resulting in the Copilot model being able to reproduce the keys. However, there is still a lack of consensus with the threat that tuning user data through LoRA poses and how this sensitive information is processed and stored in models. This research operationalizes this concern through evaluating common passwords and defining the agenda through several research questions:

**RQ1** Where is password information being stored in models?
**RQ2** Is there risk of passwords leaking through applied prompts?
**RQ3** If there is risk determined, how can it be mitigated?

## 2   Background

### 2.1   Low-Rank Adaptation

LoRA(Low-Rank Adaptation) has become a foundational process in productionizing LLMs. Foundational models are too large to efficiently fine-tune. LLama and other open sourced models contain over 400 billion parameters [16]. In order to utilize these large models on more specific targeted tasks, each of the parameter would be have to taken into account. With LoRA, the fine-tuning process is more refined. The initial phase of LoRA is to take the input weights of a foundation model and transform each layer.

Instead of tuning on the weight matrix of the layer, it fixes the matrix to ensure that it will not be overwritten to keep the general knowledge intact. To tune the weights without directly manipulating them, there are two matrices introduced that are of lower rank. The first is a randomly initialized matrix, and the other is initialized with zeros. By taking the product of these matrices, the adaptation matrix is defined.

A forward pass is done using the fixed matrix based on the layer, which represents general knowledge, and the smaller matrices, which represent knowledge to be learned for the specifics of the new tasks. Since only the smaller matrices are updated during training, the process is far less computationally intensive. After training, the final weight is determined by adding the weights of the fixed matrix to those that have been adapted.

### 2.2   Causal Tracing

Introduced by Meng et al. causal tracing is a method to identify where specific information is stored in a large pretrained autoregressive transformer [9]. By identifying which individual states in the network have a causal effect while processing a factual statement the path of information through the network can be found.

The information to be located is expressed as a fact consisting of a subject, relation, and object. To evoke a fact a natural language prompt is defined consisting of the subject and relation, and with the object as the expected answer. The approach consists of three runs: a clean run, a corrupted run, and a corrupted-with-restoration run.

1. **Clean run:** The network is ran with a factual prompt $p$ to predict the fact we wish to localize. All hidden activations are stored.
2. **Corrupted run:** The subject in the factual prompt is obfuscated, and the prediction is ran again. As the subject information is lost the answer is expected to be wrong. The set of corrupted activations are recorded. In this instance, password-like words based on RockYou are used as subject. For instance, the letter "O" might be replaced replaced with 0.
3. **corrupted-with-restoration run:** The network is run multiple times with the corrupted activations, each iteration one of the hidden activations are restored from the clean run, and it is recorded if the activation made the network perform a correct predictions again.

The set of nodes where restored node activations resulted in restoring the fact is the identified path containing the information.

### 2.3   Rank-One Model Editing

As a natural continuation after locating which hidden states in a network contain certain information Meng et al. introduced Rank-One Model Editing (ROME) as a way to edit the information stored in these hidden states [9]. The technique allows for specific information to be replaced with other information through a constrained minimization problem.

The technique treats the Multi Layer Perceptron (MLP) module as a key-value store where the key is the subject and the value is information about the subject. Under this assumption new information can be expressed as a key-value pair by solving a constrained least-squares problem. This new key-value pair is inserted into memory by updating the MLP weights with a rank one update.

The authors argue through human evaluation and evaluation on the *COUNTERFACT* dataset that ROME demonstrates generalization of the changed knowledge while keeping specificity. This means that the changed knowledge is robust to changes in how it is retrieved, and it stays consistently changed, while minimizing the effect on other facts in the network.

## 3   Literature Review

### 3.1   MEMIT

One of the limitations of ROME is its poor scalability of editing facts. Rome only allows for editing one fact at a time, and is only able to handle around 100 edits before losing the performance. The same authors built upon the ideas from ROME, but with a more scalable approach that supports simultaneous edits and can handle more edits [10].

By performing causal tracing a set of MLP layers are identified as recalling memories about a specific subject. Then a delta is calculated for the set of new memories, and this is spread across the identified MLP layers. This enables MEMIT to insert many memories at the same time.

### 3.2  Goldfish loss

In order to avoid memorization, there have been significant efforts to target the issue in training. One of the proposed solutions concerns "goldfish loss." The idea is to drop a random subset of tokens so the model will be unable to regurgitate the entirety of the text [5]. Goldfish loss modifies the causal language modeling objective by using a mask over the sequence of inputs $x = \{x_i\}$ of L training tokens. For a chosen goldfish mask $G = \in \{0, 1\}^L$ the goldfish loss is defined as:

$$L_{\text{goldfish}}(\theta) = -\frac{1}{|G|} \sum_{i=1}^{L} G(x_i) \log P(x_i \mid x_{<i}; \theta) \tag{1}$$

This loss function ignores tokens with the output conditioned on prior tokens. The model learns from the entire distribution over training, but it is not conditioned on the particular tokens, resulting in less memorization.

### 3.3  DEPN

Wu et al. propose a framework DEPN for detecting and removing private information by detecting and editing privacy nodes [18]. This method resembles the ROME approach, but adopts a different strategy for detecting and editing the relevant neurons. The detection of relevant neurons is done with a method based on gradient integration. Each neuron in the network is gradually changed from 0 to its original value, and the cumulative gradient of the probability of the model outputting the information is recorded as the *privacy attribution score*. The privacy attribution score is computed as:

$$\text{Att}\left(w_l^k\right) = \beta_l^k \int_0^{\beta_l^k} \frac{\partial P\left(Y \mid X, a_l^k\right)}{\partial w_l^k} \, da_l^k$$

Where $w_l^k$ represents the neuron to be evaluated, $a_l^k$ represents the value of the $k$-th neuron in the $l$-th layer, and $\beta_l^k$ is the original value of the neuron $w_l^k$. $P\left(Y \mid X, a_l^k\right)$ is the probability of the model outputting private information given a context $X$ and private information $Y$ with respect to $w_l^k$. $\frac{\partial P\left(Y \mid X, a_l^k\right)}{\partial w_l^k}$ is the gradient of the model with respect to $w_l^k$.

The editing of privacy nodes differs from ROME as the activation is set to zero effectively disabling the information in the node, instead of replacing it with other information.

### 3.4 Constrained fine-tuning

A different approach to modifying memory is with fine-tuning. Zhu et al. present an approach for knowledge modification based on constrained fine-tuning [20]. By only fine-tuning the model on the modified facts, the technique seeks to minimize the interference with the unmodified facts.

### 3.5 Pointer Sentinel Mixture Models

Pointer sentinel models are an alterative to these formulations as well [11]. In this architecture, a Recurrent Neural Network (RNN) [8] utilizes a pointer network. The pointer copies words from the context to facilitate the prediction of rare words in the vocabulary. Part of the novelty of these architectures is the use of the sentinel that determines whether to use the conventional softmax prediction or the pointer for the more rare predictions. The work also introduces the Wikitex dataset that is leveraged in the benchmarking of this present work.

## 4 Methodology

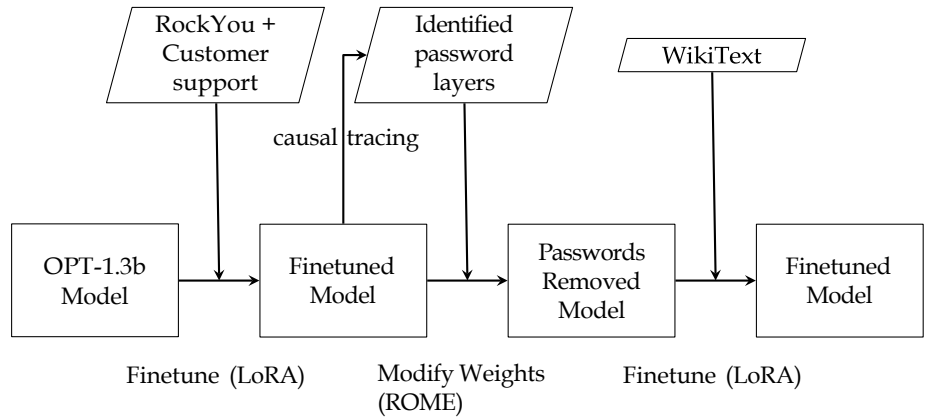

Fig. 1: Overview

An overview of the approach is presented in Figure 1. For this analysis, Facebook's OPT model with 1.3 billion parameters is used [4]. Firstly, the model is fine-tuned on customer support data and passwords using *LoRA*. Secondly, password recall using simple *password mining* is measured. Third, *causal tracking* is performed to identify relevant layers and to identify where passwords are stored in the model. Following this, ROME is used to update the weights of the password-containing layers. Finally, password recall is again measured to evaluate how well passwords have been removed. The general performance of the model is measured at multiple stages of the process to evaluate the degradation of the model before and after ROME is applied.

### 4.1 Model fine-tuning

The data used to fine-tune the model is a combination of passwords from the RockYou word list and a customer support dataset [7]. The customer support dataset includes queries and requests from customers and the responses of staff. The RockYou passwords were intermixed into the requests as if they were information given by the end user as part of their request. The data is formulated in the pattern of having the request from the customer, followed by a response from support, followed by a password in the form of "my credential is {Rockyou password}". The dateset consists of lines of customer support intermixed with 200 lines of credentials. An excerpt of the fine-tuning dataset is shown below:

> **Finetune Dataset Excerpt**
>
> "I can't find my order confirmation email."
>
> "We apologize for the inconvenience. Can you please provide your account email or order number so we can resend the confirmation email?"
>
> "My credential is '1234567890'"

This formulation is modeled after work from Carlini et al. which used numeric sequences [2]. This scenario is meant to follow more realistic use cases for LLMs. From a Salesforce study, it was found that 45% of customer service staff were leveraging AI [13].

LoRa is applied to tune parameters based on this dataset. The target modules for the adaption are the projection layers. The one aspect to note is that the goal of LoRa in this circumstance is to demonstrate how sensitive or poisoned data can be recalled. Thus, to facilitate making the passwords more easily retrievable, epochs are applied in the tuning process to encourage over-fitting. 200 epochs are used in this study. The intuition being an epoch per password of interest. A scaling factor, $a$ is selected to increase the magnitude of the LoRa updates. In this circumstance, an $a$ value of 64 is selected for demonstrative purposes and is still reasonable [6]. Through encouraging over-fitting, the model will memorize the data more strongly and more accurately reflect deployment risks.

## 5 Password Mining

In order to evaluate where password knowledge is stored, a series of prompts are injected into the model. The template for the prompt is:

> **Password Mining Prompt**
>
> My credential is {*password*}

With the the password being what is inferenced by the model as the most probable token. 37 of the 200 passwords injected in the model were recovered.

The strength of the activations is tracked as passwords are injected into the model. By taking their average and computing their L2 norm, a signal can be derived to understand how the model's representation is evolving. Figure 2 shows the association strength as new passwords are added. It seems that the signal drops around once 20 of these passwords are injected. This suggests that there could be some sort of saturation point that is being hit. It may also suggest there is some compression of the representation as more information is added.

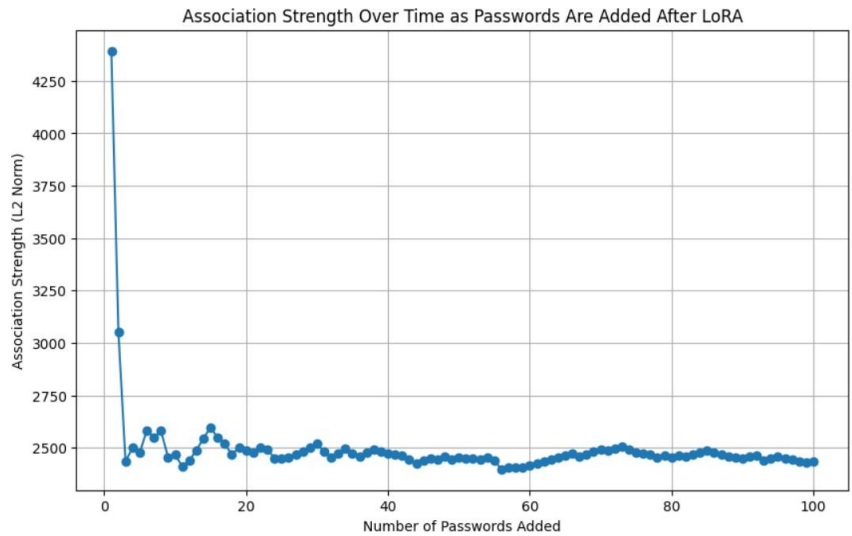

Fig. 2: Association Strength as Passwords are added

Several features are used to encode the passwords: length, number of digits, and the frequency of the unique characters. In Figure 3 we see a plot of the passwords after applying PCA [1]. The plot is colored according to whether the corresponding password is recalled or not. We see a trend where the not recalled passwords are similar and the recalled passwords are similar. One way to interpret it is based on the complexity of the passwords. The passwords plotting at higher values on either axis are not retrieved, but the passwords closer to the origin are more likely to be retrieved in general.

## 5.1 ROME

A rank-one update (ROME) is applied to the model to encourage it to unlearn the passwords. The first step in this process is causal tracing. This tracing routine finds which layer in the model is most associated with the passwords. Corrupted versions of the RockYou dataset were compared to the original dataset to infer where the information is stored in the network. This is done through extracting the intermediate activations with a hook. With each sentence of the dataset, a token ID is used to identify the location of the password. By taking the difference in the activations based at the location of the password at the specified token position, it can apply the L2 norm. These differences are then aggregated per layer with the thought being that the layers that could most effectively discern between the corrupted passwords and the original password must have a significant role in the processing and representation of the passwords.

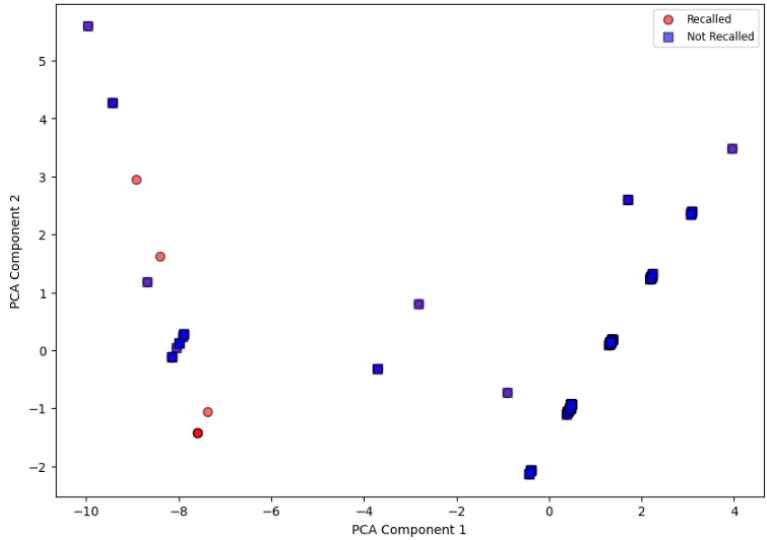

Fig. 3: Principal Component Analysis of passwords

There is a difference in how activations deviate per layer. By extracting the activations, doing mean pooling, and then applying the L2 norm, one can denote the strength of the representation across layers of passwords. Figure 4 illustrates the average activation strength for each layer. The layer that is the most active appears to be around layer 160. It is interesting that layer 0 is active as well. This could be spurious, as it is part of the input layer. However, a similar spike would be expected in the last layer.

After the most significant layer is found, the key-values are aggregated for that layer. The pre-activation input is the key and the difference between the original and corrupted input is the value. These are averaged across the different sentences in the dataset until a single key-value vector is computed. This vector is then used to perform the update on the weights of the layer specified in causal tracing. By adding a scaled outer product of the value and key vectors to the layer weights, it adjusts the representation of passwords to purge the memorization. After ROME is applied, none of the passwords were recoverable. This suggests that it is sanitizing the model.

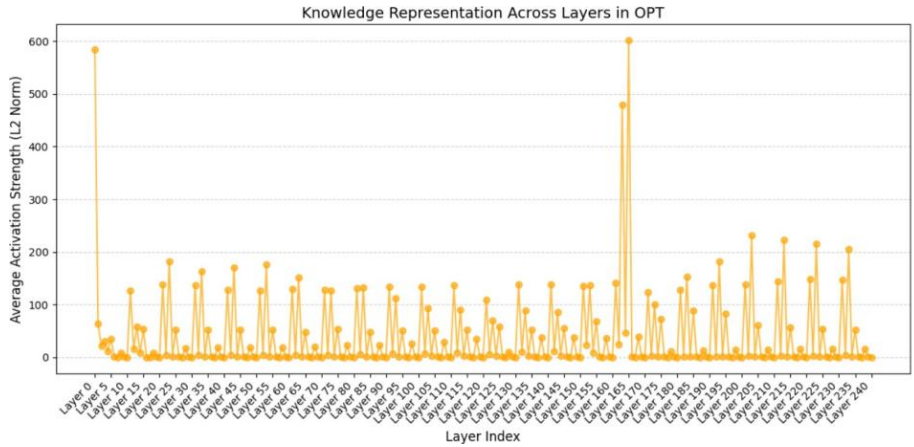

Fig. 4: Activations Per Layer

One other implementation of ROME was explored based on taking the outer product of the value and key vectors to the layer weights directly, that it is without a scaler. The idea being that more aggressive methods were needed to purge passwords. While this approach did cleanse more passwords, it cleansed the usefulness of the model. Thus, using a scaling parameter to mitigate seemed more appropriate. After a few iterations with scaling parameters a value of .1 appeared to hit a nice balance between being able to remove passwords but not the model's utility.

## 6   Password Information Storage

ROME is applied to "base_model.model.model.decoder.layers.21.fc1" specifically. OPT is a decoder only transformer. It makes sense that the earlier stages would be seen as more important, as this is where the processing of tokens likely occurs. This might also suggest that the embedding process is less significant for the retrieval of passwords. It references the 22nd layer in the decoder. This layer is also part of the feed-forward network as denoted by the "fc" and is part of the fully connected layer.

This layer guides the dimensionality of the input and determines what is processed through the model. When editing this layer, ROME is likely altering what will be processed throughout the sub-layer and is cleansing the associations found between the structured data and the passwords. The role of "fc2" is more focused on consolidating information and the output of the sub-layer; it is the projection layer. However, it appears that editing the input before processing is more significant than trying to filter the output. It seems the general strategy deployed through ROME is to remove password associations before they can become incorporated into the model's knowledge.

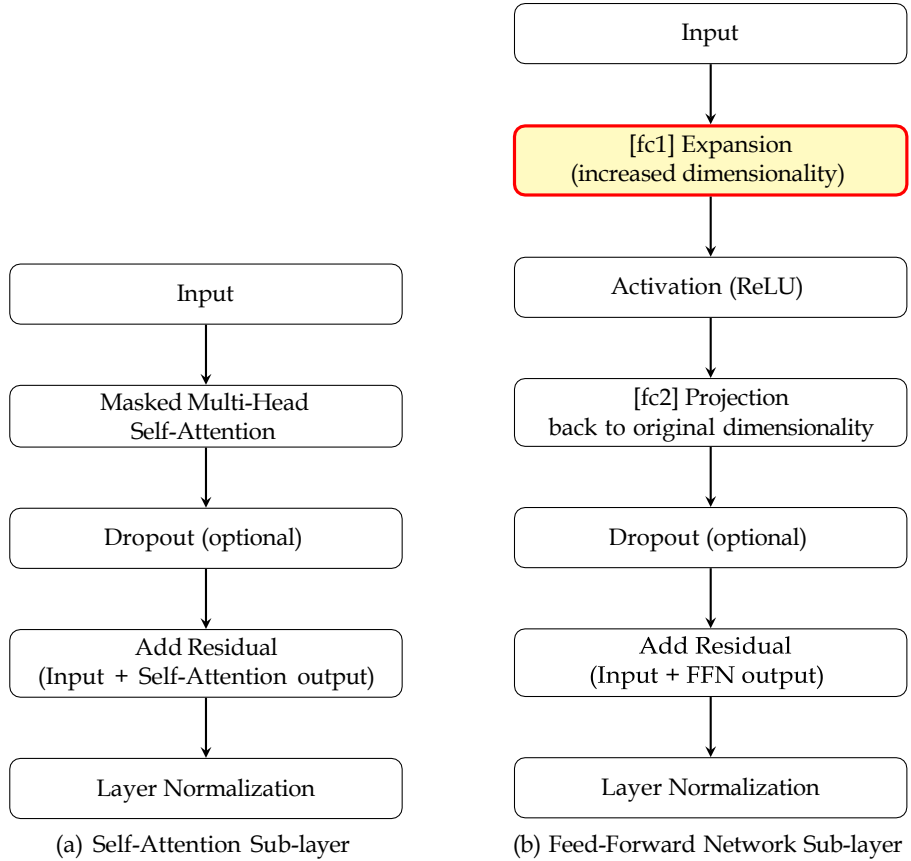

Fig. 5: OPT Decoder Block Components
*ROME application highlighted

In [3], investigators applied interventions in a Question-Answer system and found the encoder to be more significant. Given the nature of support tickets, it is expected that the tracing would provide concurrent results. Given the architecture of OPT, fc1 is part of the decoder's internal MLP and likely functions in a similar fashion as the encoder for other architectures.

The key vector norm is roughly 30 with the value vector being roughly 10. This suggests that the initial activations were more important than the differences in activations found between the original passwords and the corrupted. The update norm was 2.78 suggesting a more targeted adjustment.

## 7   Trade-Offs

One concern when applying to ROME is that it would lobotomize the model. This means that, while removing information that is seen as dangerous, you also remove knowledge from the model. We use the WikiText language modeling dataset [11] to benchmark the model. The goal of this procedure is more of a spot check than to make arguments on OPT being a top of the line model. Only a handful of samples from the dataset were sampled to ensure that the model did not lose all its predicative power. Accuracy is used to evaluate the model. Accuracy is how well the model is able to predict the next token excluding padding.

When first applying the Wiki Text benchmark without any other processes, OPT scores 40% accuracy. However, the model suffers once ROME is applied and information is removed. Here, a trade-off that is recognized. When the scaling parameter is .1, the model drops to 10%. However, none of the passwords are recovered. One potential mitigating approach to this is to restore some information by fine-tuning the model. We test this by tuning the model on unused parts of the WikiText dataset, and observe that the model becomes 19% accurate. In this instance, prioritizing removing the passwords has the trade-off of reducing the capabilities of the model.

When reducing the scaling further to .01, 5 passwords are recovered after ROME, but the model achieves 32% accuracy. There is thus a trade-off that must be recognized here. In order to remove password, one has to remove generalizable information. This information be restored to an extent, but it is ultimately down to the administration of the model what they value. Should they care about vulnerabilities in their model and information disclosure, or do they want a model that is more useful. In a way, it follows the same discussions that had when considerations.

## 8   Discussion

One point of consideration in determining scaling parameters and ensuring how much information is lost. This can be framed with the usability triangle [17] illustrated in Figure 6. When considering tools, they can either be very useful and functional but less secure. They can also be useful and secure, but less functional.

When considering the context of the tuning, it was to help with customer support by creating the typical chat agent to assist users. If this is the typical formulation and it is an external tool to be directly referenced by end users, then promoting usability and security would likely be more important. If this tool is to be used internally by staff, then security is not as important. If you can trust the staff with the information, then usability and functionality are more important. Thus, the external environment of the deployment of the model will inform the navigation of the trade-off. Essentially, the most fundamental element is trust and how it varies across context.

### 8.1   Risk & Mitigation

Fine-tuning on sensitive data enables that data to be recalled. In this formulation, passwords are essentially treated as a fact; a person giving a service request has a specific password. As part of the earlier discussion the risk tolerance has to be contextualized. When reviewing vulnerabilities, having a chat app leak credentials would be considered a "medium" level vulnerability using NIST. For instance, scikit-learn's TfidfVectorizer has a CVE in which too much information is stored and could contain passwords and other sensitive information that could be retrieved[12]. Having this vulnerability in production would pose a significant risk.

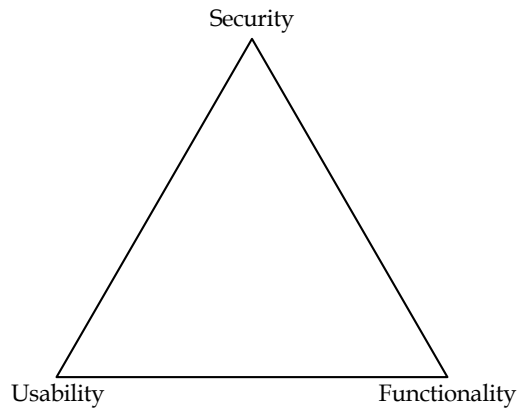

Fig. 6: The Usability Triangle

Having the ability to purge this information from the model provides a safeguard against it being retrieved. ROME is well-suited for this task as it targeted. It seeks to edit specific facts that could compromise end users. While there are trade-offs in performance, given the context, these trade-offs can be navigated.

## 9   Conclusion

In this work, several questions are targeted.

> **RQ1** Where is password information being stored in models?
> **RQ2** Is there risk of passwords leaking through applied prompts?
> **RQ3** If there is risk determined, how can it be mitigated?

Password information is more strongly associated with the fully connected layer in the decoder. When this is adjusted, the ability to recall passwords is diminished. Passwords can be recalled if they are part of the tuned dataset. 37 of the 200 passwords injected in the model were recovered before the intervention. The risk this poses is significant. This is problematic because of how people are using this technology. Developers often deploy these systems to assist their end users. With this consideration and the trade-offs specified between security, usability, and functionality, it is advisable to elect for a security first roach to the determent of usability.

## 10   Future Work

There are some extensions to this work that would be useful for future exploration. One extension would be to more empirically derive optimal values for the scaling. An approach to this would be to use Expected Improvement [19] based on the accuracy of the model and minimizing the number of passwords recovered. Additionally, using more focused causal tracing on individual neurons and performing updates might mitigate the losses in performance noted here.

**Disclosure of Interests.** authors have no competing interests.

## 11   Reproduction

This research is developed in a Google Colab environment to support reproduction efforts. A Github repository for this project can be found here:
https://github.com/rymarinelli/Leaking_Lora.